\documentclass[showpacs,noshowkeys,twocolumn,amsmath,superscriptaddress]{revtex4}
\usepackage{graphicx}
\usepackage{bbold}
\newcommand{\be}{\begin{eqnarray}}
\newcommand{\ee}{\end{eqnarray}}

\newcommand{\X}{{\bf X}}
\begin{document}

\title{Quantum Diffusive Dynamics of Macromolecular Transitions} 
\author{S. a Beccara}
\affiliation{Dipartimento di Fisica, Universit\`a degli Studi di Trento, Via Sommarive 14, Povo (Trento), I-38123 Italy.}
\affiliation{INFN, Gruppo Collegato di Trento, Via Sommarive 14, Povo (Trento), I-38123 Italy.} 
\author{G. Garberoglio}
\affiliation{Interdisciplinary Laboratory for Computational Science (LISC), FBK-CMM and Universit\`a degli Studi di Trento, 
Via Sommarive 18, I-38123 Povo, Trento, Italy}
\author{P. Faccioli~\footnote{Corresponding author: faccioli@science.unitn.it}} 
\affiliation{Dipartimento di Fisica, Universit\`a degli Studi di Trento, Via Sommarive 14, Povo (Trento), I-38123 Italy.}
\affiliation{INFN, Gruppo Collegato di Trento, Via Sommarive 14, Povo (Trento), I-38123 Italy.} 

\begin{abstract}
We study the role of quantum fluctuations of atomic nuclei in the real-time dynamics of non-equilibrium macro-molecular transitions.
To this goal we introduce an extension of the Dominant Reaction Pathways (DRP) 
formalism, in which the quantum corrections to the classical overdamped Langevin dynamics are rigorously taken into account to order~$\hbar^2$. 
We first illustrate our approach in simple cases, and compare with the results of the instanton theory.  Then we apply our method to study the $C7_{eq}\rightarrow C7_{ax}$ transition of alanine dipeptide. We find that  the inclusion of  quantum fluctuations can significantly modify the reaction mechanism for peptides. For example, the energy difference which is overcome along the most probable pathway is reduced by 
 as much as $50\%$.
   \end{abstract}
\maketitle
\noindent 

\section{Introduction}

Classical or \emph{ab-initio} molecular dynamics (MD) simulations have become a standard tool to investigate a wide range of physical systems, with  widespread  applications in chemistry, 
material science and molecular biology.
Such approaches are based on the assumption that the atomic nuclei can be treated as classical particles~\cite{QMM}. 

A classical description can be considered reliable for most atomic species comprising organic molecules and materials. 
In fact, quantum fluctuations of carbon, oxygen, nitrogen atoms at room temperature are expected to lead to small corrections.
On the other hand, quantum effects are expected to play a much more important role in the dynamics of the lightest atomic species.
For example, typical quantum fluctuations of a hydrogen atom around its equilibrium configuration in a macro-molecule at room temperature can be shown to be of the order of
fractions of the Bohr radius. 

An efficient method  was developed~\cite{Landau} to account for quantum effects in the evaluation of \emph{thermal averages}, in the semi-classical and 
 non-degenerate temperature regime 
 \be
\frac{\beta\, \hbar^2}{m} \ll \sigma^2 \qquad(\beta =1/k_B T),
\ee
where $\sigma$ is a typical length-scale characterizing the interaction between atoms.  In such an approach, averages of arbitrary configuration-dependent observables can be evaluated  to order $\hbar^2$ accuracy   by simply replacing the 
potential energy $U({\bf X})$ in the Boltzmann's weight $P({\bf X})\propto e^{-\beta U({\bf X})}$ 
with an effective semi-classical potential $U_{Q}(\X)$, which reads 
\be
\label{UQ}
U_{Q}(\X) \equiv U(\X) + \beta \sum_{i=1}^N \lambda_i  ~\vec{\nabla}_i^2 U(\X)\nonumber\\
- \frac{\beta^2}{2} \sum_{i=1}^N \lambda_i~ |\vec{\nabla}_i U(\X)|^2.
\ee
In such an expression,  ${\bf X}=(\vec{x}_1,\vec{x}_2, \ldots, \vec{x}_N)$ is a point in the $3N$-dimensional configuration space  defined by the set of all atomic coordinates  and  
\be
\lambda_i = \frac{\beta ~\hbar^2}{12 ~m_i}
\ee
 are characteristic parameters which set the length scale of quantum fluctuations of the particles. 
 For example, for a hydrogen atom at room temperature, $\lambda = 1.3 \times 10^{-4}$~nm$^{2}$, therefore $\sqrt{\lambda}$ is about 20$\%$ of the Bohr radius. 

Clearly, this approach is only useful to investigate thermodynamical properties of molecular systems. Accounting for quantum corrections to dynamics and kinetics 
is in general a much more challenging task. To this goal, a  number of methods have been proposed in the literature, such as 
centroid molecular dynamics\cite{centroid}, or instanton-based approaches\cite{instanton1, instanton1.5, instanton1.75, instanton2, instanton3}.

All of these methods represent useful tools to target specific questions.
For example, the centroid method can be used to investigate the real-time evolution of quantum many-body systems 
over short time intervals. On the other hand, it becomes 
very inefficient to investigate the long-time dynamics of thermally activated reactions. The reason is that, like  any  algorithm based on the integration of the equation of motion, it wastes most of the computational time 
to simulate the exploration of the meta-stable states, i.e. when the system is not undergoing the transition. 

Instanton-based methods provide an elegant and powerful tool to  compute 
quantum corrections to the reaction rates.  On the other hand, they do not yield direct information about the  \emph{real-time 
non-equilibrium dynamics}, since they are based on a path integral representation of the quantum partition function. 
 
In this paper, we introduce a formalism which complements the existing methods and allows to efficiently and rigorously
 investigate the real-time evolution of (macro)-~molecules in non-equilibrium 
conditions. Such a fully microscopic approach is based on the path integral representation of the solution of a Fokker-Planck equation which includes order $\hbar^2$ quantum corrections.   
In particular, our method is useful to investigate reaction mechanisms, since it yields a natural and 
unbiased reaction coordinate, and it allows to predict the time evolution of arbitrary observables during the \emph{most probable} reaction pathways. In many cases of interest, such information can be directly compared against experimental data. For example, in the context of protein folding, information about the reaction mechanism are 
available from the so-called phi-value analysis \cite{phivalue} or from single-molecule kinetic experiments --- see e.g. \cite{singlemoleculeexp} and references therein~---.  

The semi-classical extension of the DRP method we present in this work is computationally very efficient: on the one hand, it avoids wasting computational time to simulate the dynamics when the
system is not undergoing a transition to the final state. On the other hand, the inclusion of quantum corrections to order $\hbar^2$ does not involve a significant increase of the computational cost of the calculation,
since it requires to compute quantities which are already evaluated in the classical approach. 

We first illustrate our approach on a very simple two-dimensional toy-system. 
Then, we compare the effects of quantum fluctuations on the reaction pathways for H$_2$ dissociation on the Cu(110) surface
obtained in our approach ---~which holds in non-equilibrium conditions~---  and in the instanton method 
---~which applies to equilibrium conditions---.

We then perform an application to a realistic molecular transition: 
the $C7_{eq}\rightarrow C7_{ax}$ re-arrangement of the alanine dipeptide.
This is a representative example of a bio-molecular conformational reaction, where the formation and breaking of hydrogen bonds is an important driving force. While quantum fluctuations
of hydrogen atoms may play an important role in the hydrogen-bonding dynamics, they are usually neglected in standard biochemical simulations. However, as we shall see, their 
inclusion has significant effects on the reaction mechanism. 

The paper is organized as follows. In section \ref{QSEsection} we discuss the real-time quantum diffusive  dynamics of a molecular system, in the strong friction limit.
Section \ref{QDRP} represents the core of the paper, where we introduce the leading quantum corrections to the DRP approach. In section \ref{illustrative}, we illustrate our method
in a simple toy model and study the H$_2$ dissociation in Cu. In section \ref{molecule} we apply it to the alanine-dipeptide transition. 
Conclusions and perspectives are summarized in \ref{conclusions}. The mathematical details of the derivations are reported in two appendixes. 

\section{Quantum Diffusive Dynamics in the High Friction Limit}
\label{QSEsection}

Our goal is to include quantum corrections to the dynamics of systems which, in the classical limit, can be described by the overdamped Langevin dynamics. 
To this end, we consider the theory of quantum dissipative systems in the high-friction and semi-classical regime. Namely,
if $\gamma$ is a friction coefficient which describes the strength of the coupling of the system to the heat-bath, and
$\tau_0$ is the shortest time scale of the dynamics we are interested in,  we consider the long-time evolution  of the system in the limit:
\be
\label{conditions}
t\gg \tau_0 &\gg& 1/\gamma,\\
 \beta~ \gamma ~\hbar &\ll& 1.
\ee
The first inequality defines the overdamped regime. The second condition implies that quantum coherence effects are negligible, see e.g.  \cite{review} and references therein.
As a consequence, the real-time quantum dynamics in this limit  is completely specified by the diagonal part of a reduced density matrix, in which the heat-bath degrees of freedom are traced out:  
 \be
 P({\bf X},t)\equiv \langle {\bf X} | \textrm{Tr}_Y\hat \rho(t) | {\bf X} \rangle.
 \ee 
In this definition, $\hat\rho(t)$ is the time-dependent density operator of an enlarged Hamiltonian system which comprises the molecule degrees of freedom ${\bf X}$ and the heat-bath
 degrees of freedom ${\bf Y}$. The trace $\text{Tr}_Y$ is performed over the heat-bath variables ${\bf Y}$ only.
 
The equation which determines the probability density $P({\bf X},t)$ can be derived using the path integral representation of the time-dependent density matrix. 
It has been recently shown that, in the case of a one-dimensional quantum particle interacting 
 with an external potential $U(x)$,  the probability density $P(x, t)$ to leading-order in $\lambda$  obeys the Quantum Smoluchowski Equation~(QSE)~\cite{QSE, Hanggi}
\be
\label{QSE1D}
\partial_t P(x, t) &=& D \frac{d}{dx} \left( \hat{\mathcal{L}} ~P(x,t)~\right)\nonumber\\
\hat{\mathcal{L}} &\equiv& \beta \frac{d}{d x} U(x) + \frac{d}{dx} \left(1+ \lambda \beta \frac{d^2}{d x^2} U(x)\right),
\ee
where $D=1/(m \gamma \beta)$ is the classical diffusion coefficient. 
The same result was obtained also by Coffey and coworkers, in an approach based on the thermal Wigner function \cite{Coffey}.
 Notice that at finite temperature, and high-friction regime, the leading quantum corrections are independent of the friction coefficient~\cite{textbook}.

In appendix \ref{generalization} we derive the multi-dimensional generalization of such an equation to a system of $N$ atoms in contact with a heat-bath, and obtain
\be
\label{QSEq}
\partial_t P(\X, t) &=&
 \sum_{i=1}^N ~D_i~\vec{\nabla}_i \cdot
\left[
\frac{\vec{\nabla}_i (\beta~U(\X) - L_2(\X))}{1- L_1(\X)}P(\X,t )\right.
\nonumber\\
&+&\left.  \vec{\nabla}_i~\left(~  \frac{1}{1 - L_1(\X)}~ P(\X,t )~\right) \right].
\ee
$D_i=1/ (m_i \beta \gamma)$ is the  classical diffusion 
constant of the $i$-th atom of mass $m_i$, while  the functions 
\be
\label{L1}
L_1(\X) &\equiv& \beta \sum_{i=1}^N \lambda_i  ~\vec{\nabla}_i^2 U(\X)\\
\label{L2}
L_2(\X) &\equiv& \frac{\beta^2}{2} \sum_{i=1}^N \lambda_i~ |\vec{\nabla}_i U(\X)|^2
\ee
account for quantum corrections and appear also in the leading quantum correction to Boltzmann's weight ---cfr. Eq. (\ref{UQ})---.    
Indeed, it is immediate to verify that 
\be
P_{eq}({\bf X}) = \textrm{const.}~ e^{-\beta U_Q(\X)}
\ee
 is the stationary solution of Eq. (\ref{QSEq}).

In appendix  \ref{appendixB} we show that the non-equilibrium probability density $P(\X,t)$ which solves  Eq. (\ref{QSEq}) can in principle be sampled by integrating an associated  quantum Langevin Eq. (QLE) with a multiplicative noise~\cite{QSE}:
\be
\label{QLEi}
\vec{\dot x}_i &=& D_i  \beta \left( -\vec{\nabla}_i~U(\X) + \vec{Q}_i(\X)\right)
\nonumber\\ && 
+ \sqrt{ 2 D_i (1+L_1(\X)) }~\vec{\xi}_i(t),
\ee 
where $\vec{\xi}_i(t)$ are delta-correlated white Gaussian noises, with unit variance
and $\vec{Q}_i(x)$ are effective ``quantum forces", whose definition depends on the choice of the stochastic calculus. In particular, in the so-called Ito calculus $\vec{Q}_i$ reads
\be
\vec{Q}^{\textrm{Ito}}_i(\X)  = \frac{1}{\beta} \vec{\nabla}_i L_2(\X) -  L_1(\X)~\vec{\nabla}_i U(\X).
\ee

In order to investigate the effect of the quantum corrections to the Langevin dynamics, it is instructive to consider the diffusion close to potential energy extrema, in local harmonic approximation.
In this limit, the quantum Langevin Eq. (\ref{QLEi}) in the Ito calculus reads
\be
\label{LHa}
\dot {\bf X} &=& - \frac{1}{m \gamma}\left[\hat{\mathbb{1}} +  \lambda \beta ((\textrm{Tr}\hat{\mathcal{H}}_0)\hat{\mathbb{1}} - \hat{\mathcal{H}}_0)\right]~ \hat{\mathcal{H}}_0~({\bf X}-\X_0)\nonumber\\
 &+& \sqrt{\frac{2}{\beta  m \gamma} \left[\hat{\mathbb{1}}  +  \lambda \beta \textrm{Tr}\hat{\mathcal{H}}_0 \right]}~ {\bf \eta}(t)\nonumber\\
\ee  
where $\hat{\mathcal{H}}_0$ is the Hessian matrix at the extremum configuration ${\bf X}_0$. For sake of simplicity
we have assumed that each degree of freedom is characterized by the same quantum parameter $\lambda$.

From Eq. (\ref{LHa}) it follows that, near the extrema of the potential energy surface, the quantum contribution to the diffusion coefficient which multiplies the random force $\eta(t)$
can be re-absorbed by a rescaling of the thermal energy. 
\be
\label{rescaling}
\frac{1}{\beta}  \rightarrow \frac{1}{\beta'} = \frac{1}{\beta}~ (1+  \lambda \beta \textrm{Tr}\hat{\mathcal{H}}_0).
\ee 

\begin{figure}[t]
  \includegraphics[width=9 cm]{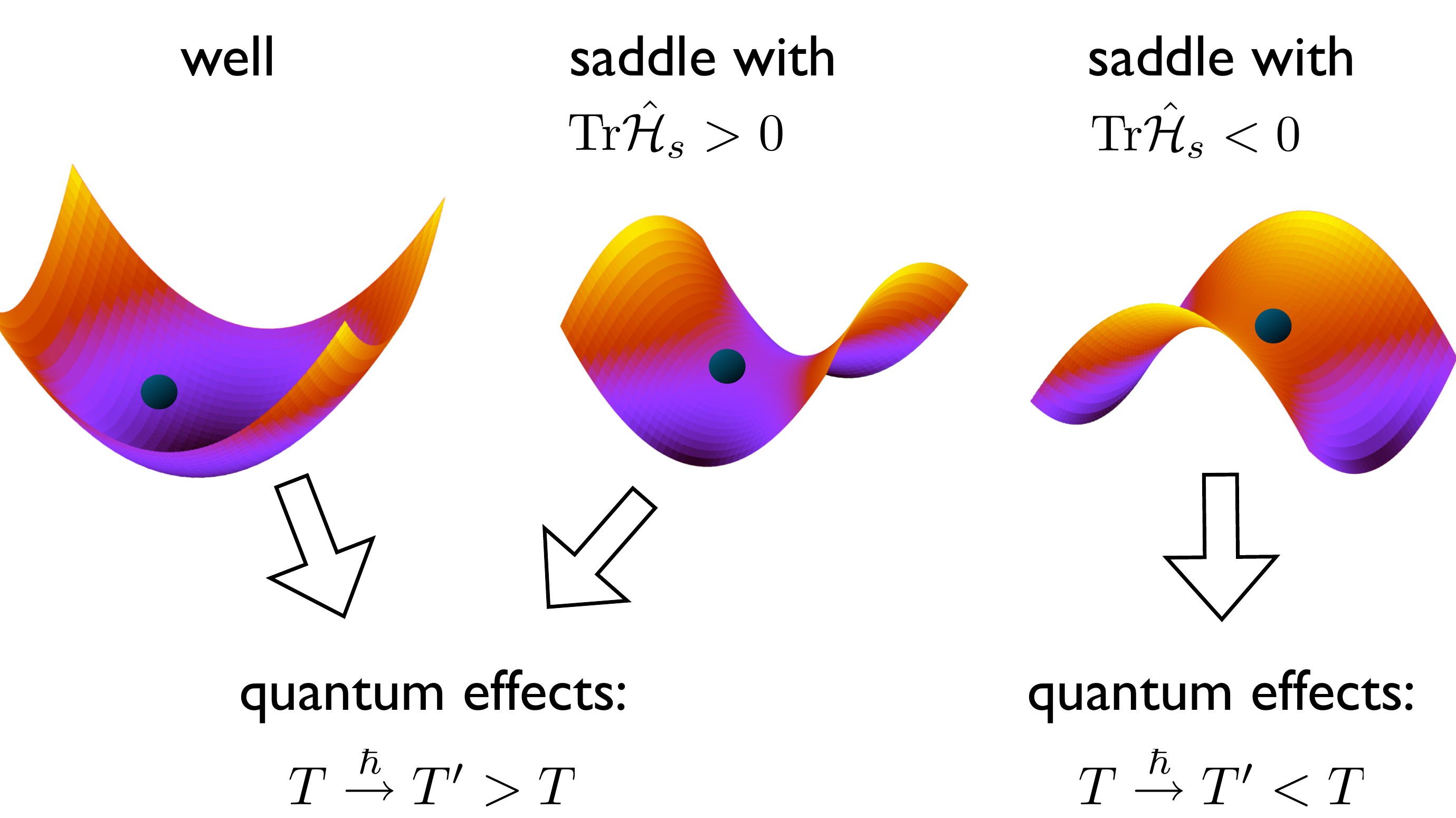}
  \caption{Interpretation of the quantum corrections to the stochastic diffusion: in the stable wells  quantum fluctuations lead to an increase of the diffusion constant, hence the quantum diffusive motion is qualitatively
  similar to a classical one at a higher temperature. However, in the vicinity of  a saddle where $\textrm{Tr}\hat{ \mathcal{H}}_0<0$, quantum effects reduce the diffusion constant. Hence, the
   quantum diffusive motion is qualitatively analog to a classical one, at a lower temperature. }
  \label{Fig1}
  \end{figure}

In particular, close to a minimum of the potential energy surface one has $\textrm{Tr}\hat{\mathcal{H}}_0 \simeq \textrm{const.}> 0$, 
hence  the quantum system diffuses like a classical one in which the heat-bath has a \emph{higher} temperature (see left panel of Fig. \ref{Fig1}).
On the other hand, near the saddles of $U(\X)$, i.e. in the transition regions, the 
Hessian matrix is not positive definite and its trace can become negative. 
In this case, the quantum system diffuses like a classical one in which the heat-bath has  a  \emph{lower} temperature (see right panel of Fig. \ref{Fig1}).

We note that, in the particular case of one-dimensional systems, the quantum force  $\vec{Q}^{\textrm{Ito}}_i(\X)$  in local harmonic approximation vanishes identically. Hence,  near the extrema  of the potential energy, the entire $o(\lambda)$ 
correction to the one-dimensional Langevin dynamics is equivalent to a rescaling of the temperature. 
In higher dimensional systems this is in general no longer the case, since the $o(\lambda)$ correction to the force  is not identically null.  

We also note  that the effective lowering of the temperature induced by the quantum effects may hardly affect the analysis of thermodynamical quantities, since the transition regions give in general small contributions to equilibrium averages. On the other hand, it may have an important effects on non-equilibrium reactive trajectories, which by definition cross the transition region. 
 
In practice, for typical molecular systems,  the direct integration of the QLE (\ref{QLEi})  can only be used for investigating very fast processes, or small thermal fluctuations around the local equilibrium configurations.  
On the other hand, for most molecular systems,  integrating such an equation of motion to investigate the long-time dynamics of a rare activated transition would be  computationally extremely expensive. 
In the next section we discuss how this difficulty can be  rigorously overcome in the DRP approach.

\section{Quantum Corrections to the Dominant Reaction Pathways}  
\label{QDRP}

The DRP approach was originally developed to  study the dynamics of rare thermally activated transitions in  systems  obeying the classical Langevin equation~\cite{DRPtheory1, DRPtheory2,
DRPrate, Elber0}. 
A remarkable advantage of  the DRP approach with respect to the MD algorithm is that the computational cost of determining the most probable pathways in a rare thermally activated transition depends neither on the height of the free energy barrier which must be overcome, nor on the existence of  gaps in the time scales associated with the system's dynamics.

The DRP method has been tested so far on conformational reactions of toy-models~\cite{DRPtheory2} and biomolecular systems~\cite{DRPtest1, DRPtest2, DRPtest3}. 
The same approach  was then applied to investigate \emph{ab-initio} both chemical reactions~\cite{QDRP1} and the folding of a peptide chain~\cite{QDRP2}. 
In these two simulations, the molecular energy $U(\X)$ and its first and second derivatives were obtained directly from the calculation of the ground state electronic structure, without resorting to empirical force fields.

We now extend the DRP formalism to account for quantum corrections in the Langevin dynamics of the atomic nuclei. 
In appendix \ref{appendixB} we derive the path integral representation of the solution of the QSE (\ref{QSEq}):
\be
\label{PI2}
P(\X_f,t|\X_i) &=&   \mathcal{N}(\X_f, \X_i)~\int_{\X_i}^{\X_f} \mathcal{D}\X~ e^{-(S_{eff}[\X] + S_{eff}^Q[\X])}.\nonumber\\
\ee
the factor $\mathcal{N}(\X_f, \X_i)$ is defined in the appendix \ref{appendixB} and does not affect the relative statistical weight of the reaction paths. The functional $S_{eff}[\X]$ is called the (classical) 
effective action and is given by  
\be
\label{Seff}
S_{eff}[\X] &\equiv& \int_0^t d\tau~\left( \sum_{i=1}^N \frac{\dot{\vec x}_i^2(\tau)}{4 D_i} + V_{eff}[\X(\tau)]~\right),\qquad
\ee
with
\be
\label{Veff}
V_{eff}(\X) &\equiv& \sum_{i=1}^N \frac{D_i \beta^2}{4} \left(|\nabla_i U(\X)|^2 - \frac{2}{\beta} \nabla_i^2 U(\X) \right).\qquad
\ee

Quantum effects are taken into account through the term 
\be
S_{eff}^Q[\X] = \int_0^t d \tau ~V_{eff}^Q[\X(\tau)],
\ee
where
\be
\label{VQeff}
&&V_{eff}^Q(\X) = \sum_{i=1}^N \frac{D_i}{4} \beta^2 |\vec{\nabla}_i U(\X)|^2 L_1(\X) \nonumber\\
&& +  \frac{1}{2} \beta D_i \vec{\nabla}_i \cdot \vec{Q}_i(\X) + \frac{1}{2} \beta D_i  \vec{\nabla}_i L_1(\X) \cdot \vec{\nabla}_i U(\X). \quad
\ee
$V_{eff}(\X)$ and $V_{eff}^Q(\X)$ will be referred to as the classical and the quantum component of the effective potential, respectively. Note that they depend on the molecular energy $U(\X)$, on the 
viscosity, and on the temperature of the heat-bath. 

For large systems, evaluating the quantum part of  the effective potential may be quite computationally expensive, since this term contains a summation over derivates of the potential energy up to fourth order.  
A  reduction of the  computational cost can be obtained by restricting the summation in the quantum terms $L_1(\X)$ and $L_2(\X)$ to the hydrogen atoms only. This is a good approximation,  
since the quantum constants  $\lambda_i$ of the heavier atoms in biomolecules is at least one order of magnitude smaller.
In addition, if the reaction under investigation is thermally activated, the potential energy barriers which must be overcome are much larger than the average thermal energy $1/\beta$. Thus in this case, solely the leading terms in the expansion of  $V_{eff}^Q(x)$ 
in powers of $1/\beta$ may  be retained:
\be
\label{leadingVQ}
V_{eff}^Q(\X) &\simeq& \sum_{i=1}^N  \beta^2 D_i \left(\frac{1}{4} |\vec{\nabla}_i U(\X)|^2 L_1(\X) + \ldots\right), \qquad
\ee
where the dots denote the sub-leading terms in $1/\beta$. Hence, the leading term in the quantum component of the effective potential only involves  the first and second derivatives of the potential energy. 
Since these terms already appear in the classical effective potential, evaluating the quantum corrections in this limit does not appreciably increase the computational cost of the calculation. 

From this point on, the derivation of the DRP formalism with quantum effects is completely analogous to the classical DRP approach: the integrand in Eq. (\ref{PI2}) expresses the statistical weight of the path connecting the initial and final configurations, in a time interval $t$. The exponents $\exp(-S_{eff}[\X])$ and $\exp(-S_{eff}^Q[\X])$ represent   the classical and quantum contributions to the probability of a given path, respectively. In particular, the most probable (or dominant) reaction pathways are those which minimize the total effective action $S[\X]= S_{eff}[\X]+S_{eff}^Q[\X]$, and
are a solution of the equation of motion

\be
\label{DRPeom}
\frac{1}{2 D_i} \ddot {\vec{x}}_i = \vec{\nabla}_i(V_{eff}(\X)+ V^Q_{eff}(\X))
\ee
with boundary conditions 
\be
\label{bc}
\X(t) &=& \X_f,\nonumber\\
 \X(0) &=& \X_i. 
\ee

The numerical advantage of the DRP  approach follows from observing that the equation of motion for the dominant paths  conserves an effective energy.
 In particular,  Eq. (\ref{DRPeom}) conserves the quantity
\be
\label{Eff}
E_{eff}= \sum_{i=1}^N \frac {1}{4 D_i} \dot {  \vec{x}}^2_i(t)  - [~V_{eff} \left( \X(t) \right) + V^Q_{eff} \left( \X(t) \right) ~].
\ee
This property makes it possible to switch from the {\it time}-dependent Newtonian description to the {\it energy}-dependent Hamilton-Jacobi (HJ) description. 
To this goal, it is convenient to introduce the rescaled atomic coordinates 
\be
\vec{y}_i \equiv \frac{1}{\chi_i}~\vec{x}_i,
\ee where 
$\chi_i \equiv \sqrt{D_0/D_i}$ is a dimensionless scaling parameter and $D_0$ is an arbitrary reference diffusion coefficient, which has been introduced to ensure the correct dimensionality of the $\vec{y}_i$ variables. 

Based on this definition, the solutions of Eq. (\ref{DRPeom}) with the appropriate boundary-conditions are  obtained by minimizing the functional 
\be
\label{SHJ}
&&S_{HJ}[{\bf Y}]=\sqrt{\frac{1}{D_0}}\int_{{\bf Y}_i}^{{\bf Y}_f}dl\left\{E_{eff} ~+ \right. \nonumber\\
&&\left.  V_{eff}\left[\chi_1 \vec{\bar{y}}_1(l),..,
\chi_N \vec{\bar{y}}_N(l)\right] + \right.\nonumber\\
&& \left. V^Q_{eff}\left[\chi_1 \vec{\bar{y}}_1(l),..,
\chi_N \vec{\bar{y}}_N(l)\right]\right\}^{1/2}\nonumber\\
\ee
where $dl= \sqrt{ \sum_{i=1}^{N} d \vec{y}_i^2}$. Notice that, since $dl \propto \sqrt{D_0}$, $\vec{y}_i \propto 1/\sqrt{D_0}$ and $\chi_i\propto \sqrt{D_0}$, the arbitrary reference diffusion coefficient $D_0$ cancels out
in the effective action $S_{HJ}[x]$.

In the HJ effective action (\ref{SHJ}) the time variable has been replaced by the curvilinear abscissa $l$, which has the dimension of a length. The crucial point is that in molecular systems there is no decoupling of the intrinsic length scales. As a result,  in order to describe reactions as complex as a conformational transition of a peptide chain, only about  
$100$ fixed $dl$  steps are usually sufficient to reach a  convergent discretized representation of the integral in Eq.~(\ref{SHJ}). This number should be compared 
with the $10^9-10^{12}$ MD time steps required to simulate a single protein folding transition with mean first-passage time in the $\mu$s~--~ms range.

In the DRP formalism, it is possible to recover the information about the real-time evolution of the system. In fact the time at which a given configuration of the most-probable path is visited is given by
the equation:
\be
\label{time}
&&t({\bf Y}) = \sqrt{\frac{1}{4 D_0}}  \int_{\bf Y_i}^{{\bf Y}} dl\left\{E_{eff} +V_{eff}\left[\chi_1 \vec{\bar{y}}_1(l),..,
\chi_N \vec{\bar{y}}_N(l)\right] \right. \nonumber\\
&&\left. + V^Q_{eff}\left[\chi_1 \vec{\bar{y}}_1(l),..,
\chi_N \vec{\bar{y}}_N(l)\right]\right\}^{-1/2}.
\ee
Notice that also the transition time does not  depend on the specific choice of the reference diffusion parameter $D_0$, as expected.

The total time is determined by the choice of the effective energy parameter $E_{eff}$. Its numerical value should not be chosen unrealistically large, 
to avoid introducing a bias towards ultra-fast transitions. 

In practice, finding the dominant reaction pathway amounts to minimizing a discretized version of the effective HJ functional:
\begin{equation} \label{effact_discr}
 S_{HJ}^{d} [{\bf Y}]= \sum_{m=1}^{N_s-1} \sqrt{ \frac{1}{D_0} \left[
E_{eff} + V_{eff}\left( {\bf Y}_m \right) \right] } \; \Delta l_{m,
m+1}
\end{equation}
where $\Delta l_{i, i+1}$ is the Euclidean distance between the slices $i$ and $i+1$, i.e
$
  \Delta l_{i, i+1} = \sqrt{ \left| {\bf Y}_{i+1} - {\bf Y}_{i} \right| ^2 }.
$

In the discretized representation of the HJ effective action
(\ref{effact_discr}), the width of the distribution of the Euclidean
distances between consecutive path slices, $\Delta l_{i,i+1}$, should
not be allowed to increase in an uncontrolled way, in order to prevent
all frames to collapse into the reactant or product configurations. As
discussed in \cite{QDRP1, QDRP2}, the most convenient way to achieve this is
to introduce a Lagrange multiplier in the minimization algorithm, which holds fixed the
ratio between the mean-square deviation from the average of the
inter-slice distances $\sigma^2$ of the average square inter-slice distance $\langle \Delta l^2 \rangle$.

\section{Illustrative Test Examples}
\label{illustrative}

It is instructive to illustrate the quantum version of the DRP approach in simple systems, before tackling realistic molecular transitions.  

\subsection{A two-dimensional toy model}

First, we consider  the diffusion of a point particle  in the two-dimensional potential 
\be
\label{Utoy}
U(x,y)= \sum_{i=1}^{3} A_i \exp [ -\alpha_i ( x - \bar x_i )^2 - \beta_i ( y - \bar y_i )^2 ]
\ee
The parameters of the potential  are given in table \ref{tabletoy}. The temperature of the heat-bath was set to $300$~K, the mass of the particle was chosen to be $m=1$~u.

\begin{table}
\begin{tabular}{|c |c c c c c|}
\hline
\# of Gaussian & $A_i$ [eV]  & $\alpha_i$ [\AA$^{-2}$] & $\beta_i $ [\AA$^{-2}$] & $\bar x_i$ [\AA] & $\bar y_i$ [\AA] \\
\hline
1 & 1 & 3.5 & 7 & 0 & 0 \\
2 & 1 & 3.5 & 7 & 0 & 2 \\
3 & 2 & 5 & 10 & 0 & 1\\
\hline
\end{tabular}
\caption{The parameters specifying the two-dimensional energy surface of the toy model defined in Eq. (\ref{Utoy}). }
\label{tabletoy}
\end{table}
 
\begin{figure}[t]
\includegraphics[clip=,width=8.5  cm]{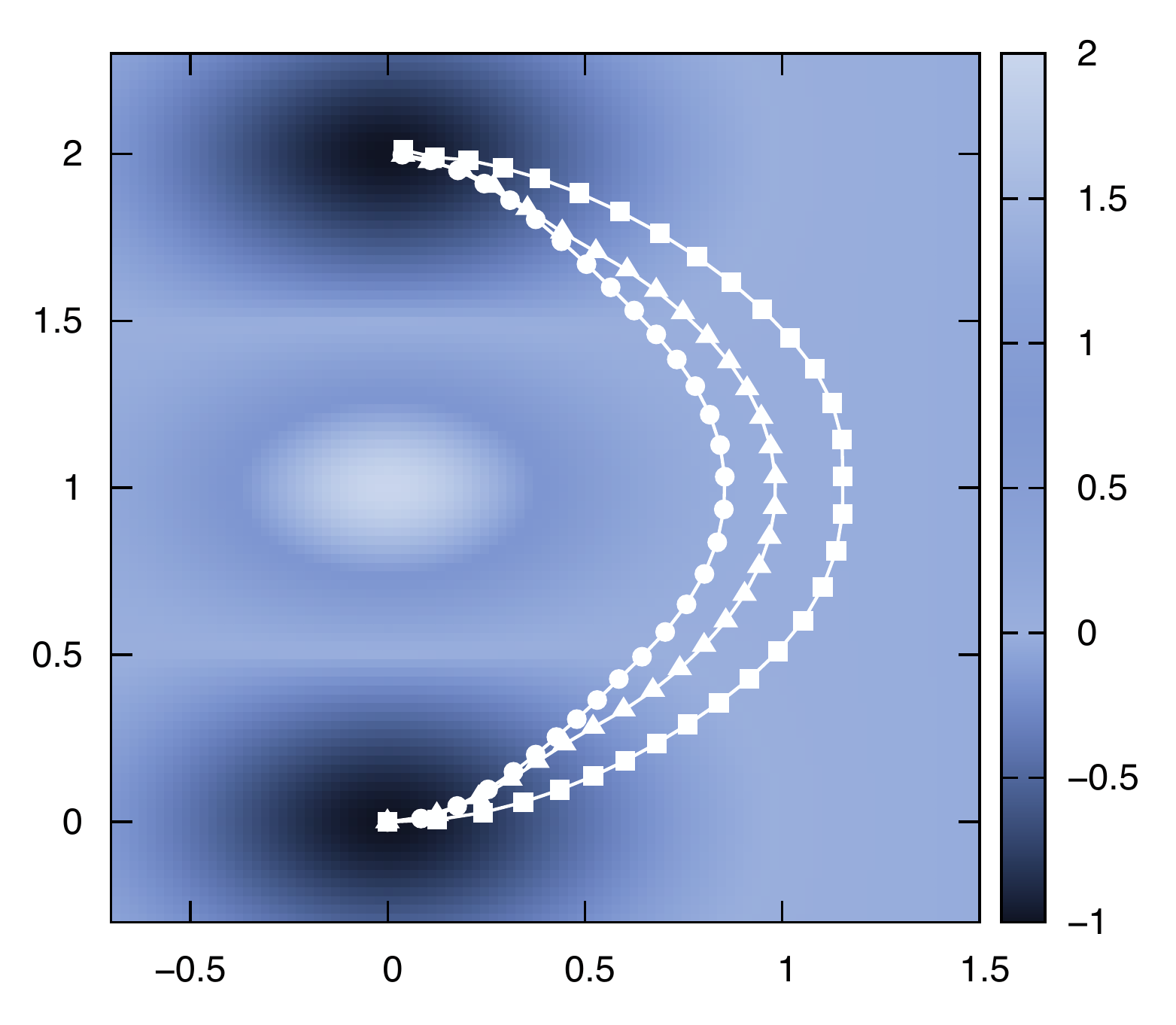}
\caption{The dominant reaction pathway in classical Langevin dynamics in the toy model defined by the potential energy (\ref{Utoy}). The circles denote the classical dominant reaction pathway, obtained minimizing
the HJ function (\ref{SHJ}),  the squares represent 
the minimum-energy path obtained from (\ref{SMEP}) and the triangles the dominant pathway with quantum corrections, obtained including the quantum component of the effective potential in the HJ action. }
\label{Fig2}
\end{figure}

In Fig. \ref{Fig2} we compare the dominant reaction pathway obtained in the classical DRP approach  (circles), with the one computed keeping into account  
the quantum correction  (triangles). In addition we plot the minimum-energy path (squares), obtained by minimizing the functional
\be
\label{SMEP}
S_{MEP} = \int_{\X_i}^{\X_f} dl \sqrt{|\nabla U[\X(l)]|^2}
\ee 
which corresponds to the  classical dominant path in the low-temperature and long transition-time limit \cite{Elber0}.  In the background we plot  the energy map.
We observe that the quantum corrections on the dominant path are appreciable. We find that, in this case,  it tends to approach the minimum-energy path  in the transition region.

\subsection{H$_2$ dissociation on the Cu(110) surface}

The DRP formalism allows to compute the quantum corrections to the real-time dynamics of diffusion-driven reactions in non-equilibrium conditions, i.e. for time intervals 
much smaller than the thermal relaxation time. Quantum effects on 
reaction kinetics have also been studied in the context of instanton-based approaches~\cite{instanton1, instanton1.5, instanton1.75, instanton2, instanton3}. Such  methods are mostly used to 
evaluate  the reaction rates and are based on the saddle-point expansion of the imaginary-time path integral (i.e. the quantum partition function). 
The corresponding saddle-point paths (i.e. the instantons)
do not directly relate to physical trajectories, hence to the real-time dynamics of the system. However, they provide information about the most often visited configurations in the transition region
at thermal equilibrium, and therefore have been used to evaluate the change in the free-energy barrier due to quantum effects.

It is reasonable to expect that the leading quantum correction to the free-energy barrier should be qualitatively consistent with the leading quantum correction
 to the energy barrier overcome by the most probable reaction pathways.
   Hence, it is interesting to compare the results obtained in the DRP  and instanton approaches. To this end, we consider the  H$_2$ dissociation on the Cu(110) surface, a reaction which has been 
investigated in detail, using instanton methods~\cite{instanton1.5,instanton1.75, instanton2}. 
In these studies it was shown that the quantum corrections lead to an effective reduction of the free-energy barrier with respect to a classical calculation. 
In particular, at $300~$K the quantum corrections lower the free-energy barrier by about $0.1$~eV~\cite{instanton1.75}.  
 \begin{figure}[t]
\includegraphics[clip=,width=8.5  cm]{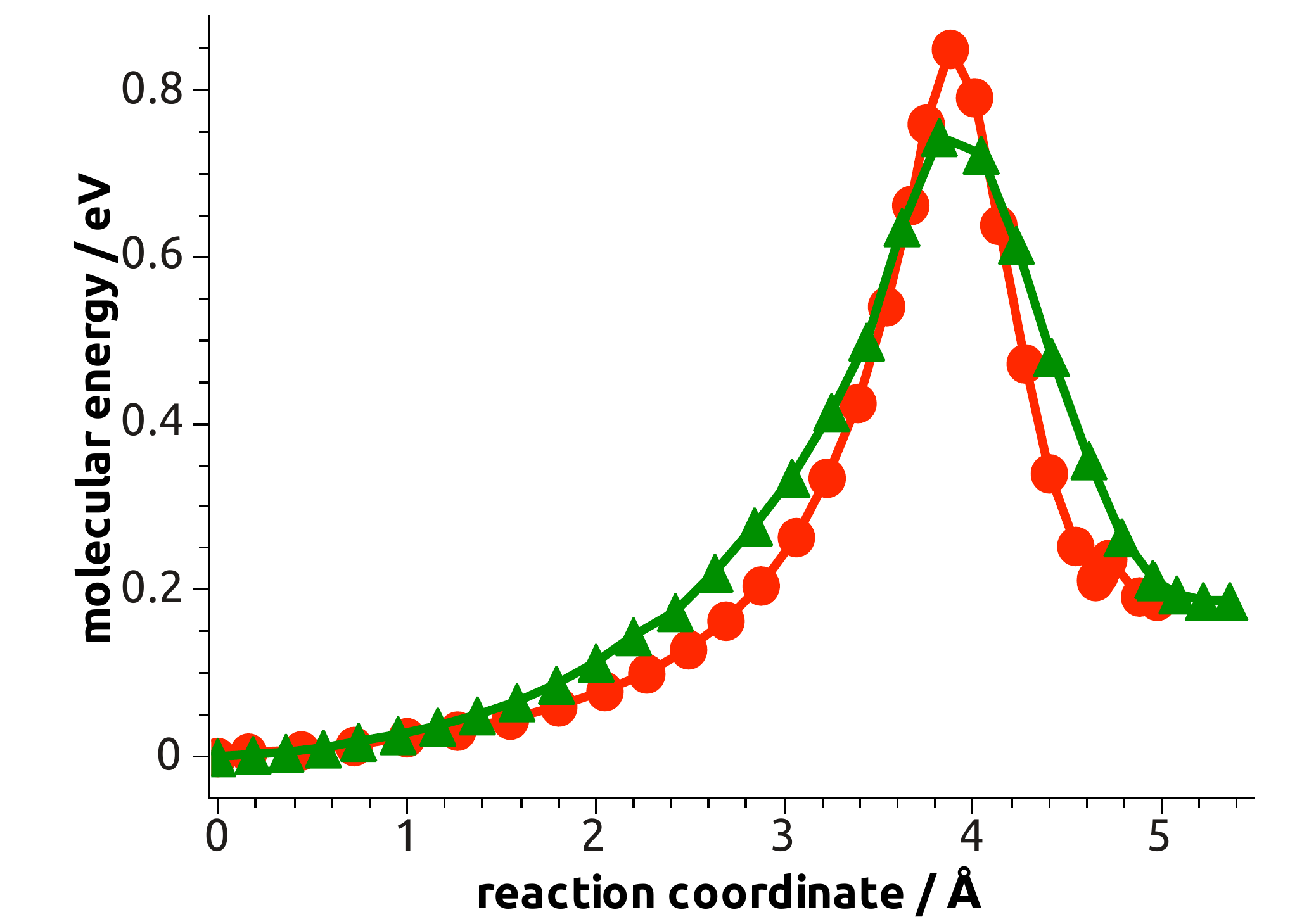}
\caption{The molecular energy evaluated along the classical (circles) and quantum (triangles) dominant reaction pathways, for  the H$_2$ dissociation on the Cu(110) surface, at 300~K.}
\label{Fig3}
\end{figure}

We have used the DRP approach to study the same reaction, adopting the same interaction potential (defined in detail in Ref.~\cite{instanton2}).
We have calculated classical and quantum dominant reaction
 paths and used them to evaluate the quantum correction to the molecular energy barrier overcome along the reaction path. We found that at $300$K quantum effects lower such a 
 barrier by about $0.1$~eV, which is compatible with the free-energy change calculated using the instanton method --- see  Fig.~\ref{Fig3}---. 
 
\begin{figure}[t]
\includegraphics[clip=,width=8.5  cm]{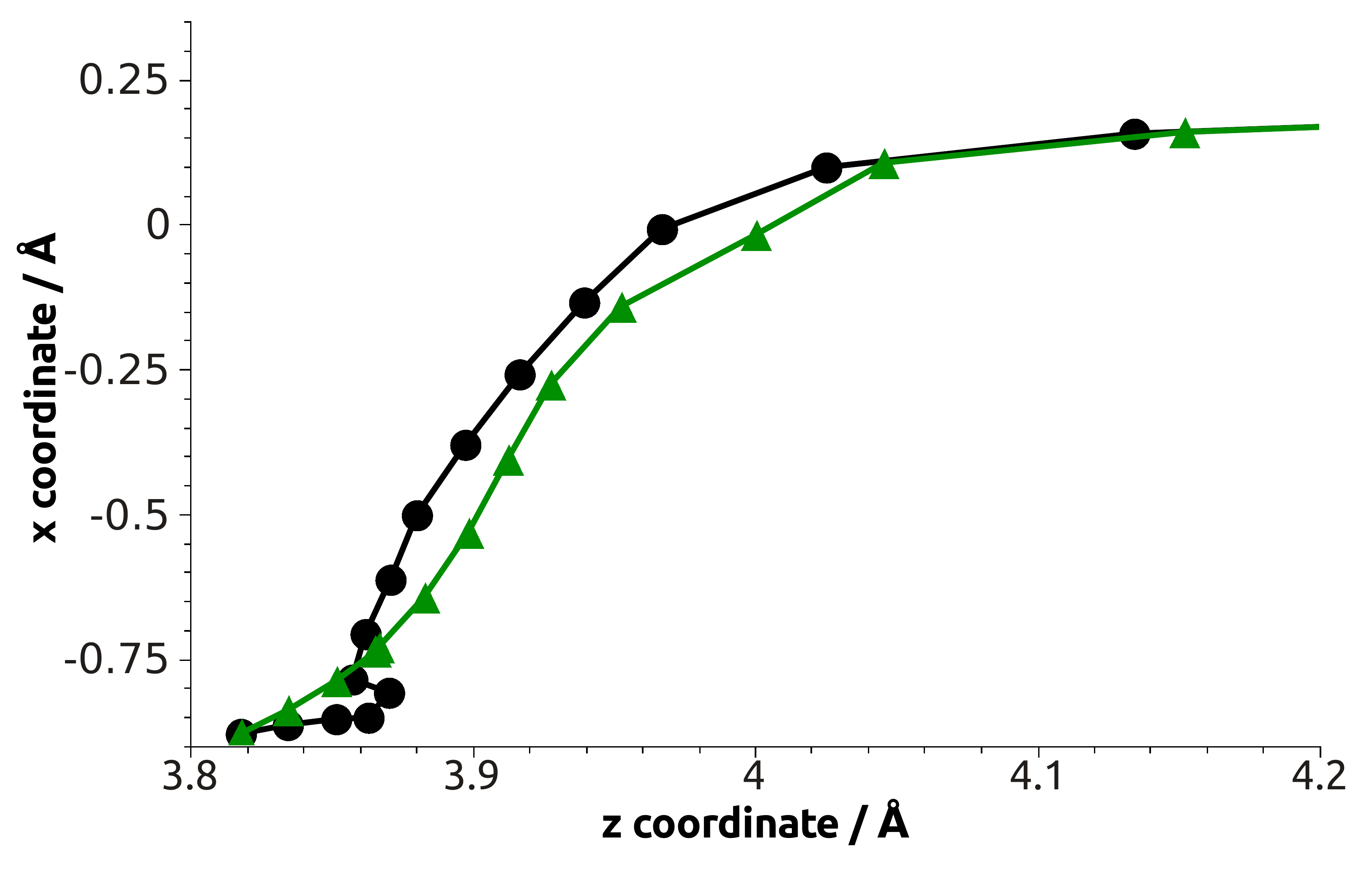}
\caption{The reaction pathway for  the H$_2$ dissociation on the Cu(110) surface, projected onto the plane selected by the $x$ and $y$ coordinates of one of the two hydrogens.
 The circles denote the
 minimum-energy  path and the triangle represent the quantum dominant  path. }
\label{Fig4}
\end{figure}

A further qualitative insight on the effects of quantum corrections can be inferred by  comparing the quantum dominant reaction pathway and the instanton trajectories calculated in \cite{instanton2}.
The DRP result is shown in Fig.~\ref{Fig4} where it is compared to the minimum energy path. We see that the dominant reaction path is shorter than the MEP,  again qualitatively agreeing
with the results shown in Fig.~3 of Ref. \cite{instanton2}.  

\section{Conformational Transition of a Peptide}
\label{molecule}

We now apply the same method  to investigate the role of quantum fluctuations in a prototypical bio-molecular conformational reaction, namely the $C7_{eq}\rightarrow C7_{ax}$ transition of the alanine dipeptide.

Let us begin by showing that the semiclassical approximation which underlies the present approach is amenable to investigating the 
conformational dynamics of a peptide. 
To this end, we observe that the typical diffusion coefficient for an amino acid of mass $m=80$~u in water is $D = \frac{1}{m \beta \gamma} \simeq 1.2 \times 10^{-3}$
 nm$^2$ ps$^{-1}$, hence $\gamma \simeq 6$~ps$^{-1}$. The condition for a semiclassical treatment of the dynamics is therefore realized: $\gamma\, \beta\, \hbar \simeq 0.2$.
The smallest time scale we can reliably describe using the overdamped limit is of the order of $1/\gamma\simeq 0.2$~ps.

The molecular energy was obtained from the Assisted Model Building with Energy Refinement
(AMBER99) empirical force field \cite{AMBER}, without solvent-induced interactions, at a temperature of $25^o$~C. 
The initial and final configurations of the peptide were obtained by minimizing the potential energy.

 For realistic reactions,  the global minimization of the HJ effective action is in general a challenging task. 
The main difficulties arise from the  ruggedness  of the effective potential and the high dimensionality of molecular systems. As a result, the most commonly used global optimization 
algorithms --- such as e.g. simulated annealing--- tend to get stuck in secondary minima of the action functional. Clearly, in this case,
 the calculated dominant paths would be strongly biased by the choice of the initial trial path. 

Our previous tests on molecular reactions have shown that the Fast Inertial Relaxation Engine (FIRE) method ~\cite{FIRE}  offers a good compromise between performance and 
simplicity~\cite{QDRP1}. The FIRE algorithm is based on a modified dynamics approach and is less prone to remain stuck in local minima 
than other minimization procedures such as conjugate gradients or Broyden--Fletcher--Goldfarb--Shanno methods. 
Moreover, we found that the FIRE algorithm was more efficient than other global methods like for instance simulated annealing. 

\begin{figure}[t]
\includegraphics[width=9.5 cm]{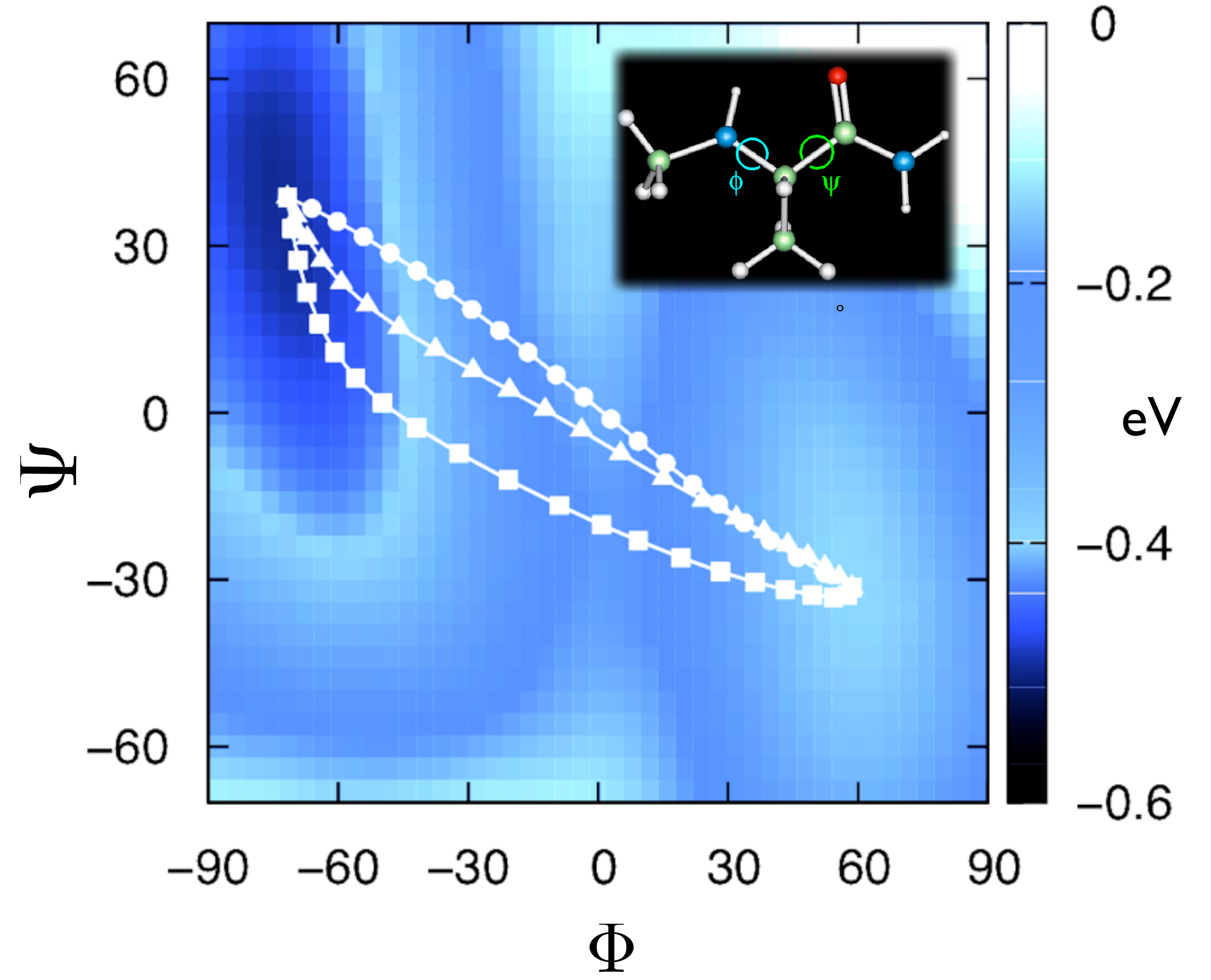}
\caption{The structure of the dominant reaction pathways for the the $C7_{eq}\rightarrow C7_{ax}$ transition of alanine dipeptide, obtained in different approaches
and projected on the Ramachandran plane.  In the background is reported the free energy landscape obtained in the classical approach.}
\label{Fig5}
\end{figure}
The minimization protocol  adopted in the present work was  the following: we generated an initial trial path consisting  of a linear trajectory which connects the initial and final points
in the Ramachandran plane specified by the $\psi$ and $\phi$ dihedral angles of the di-peptide \cite{QMM}. Such a path was discretized using 100 equally-displaced frames.
The path so obtained was initially relaxed the by means of a Nudged Elastic Band (NEB)~\cite{NEB} minimization. 
This step is crucial in order to avoid instabilities in the subsequent DRP minimization algorithm.
The NEB path was then used as a starting point for the minimization of the classical DRP action, followed by the minimization of the complete quantum DRP action.
The effective energy parameter $E_{eff}$ was chosen to be $10\%$ larger than the maximum value of $|V_{eff}[x]|$ along the NEB path.    This condition ensures a 
long transition time, and avoids that, during the minimization, the HJ effective action  becomes complex.  

\begin{figure}[t]
 \includegraphics[width=8 cm]{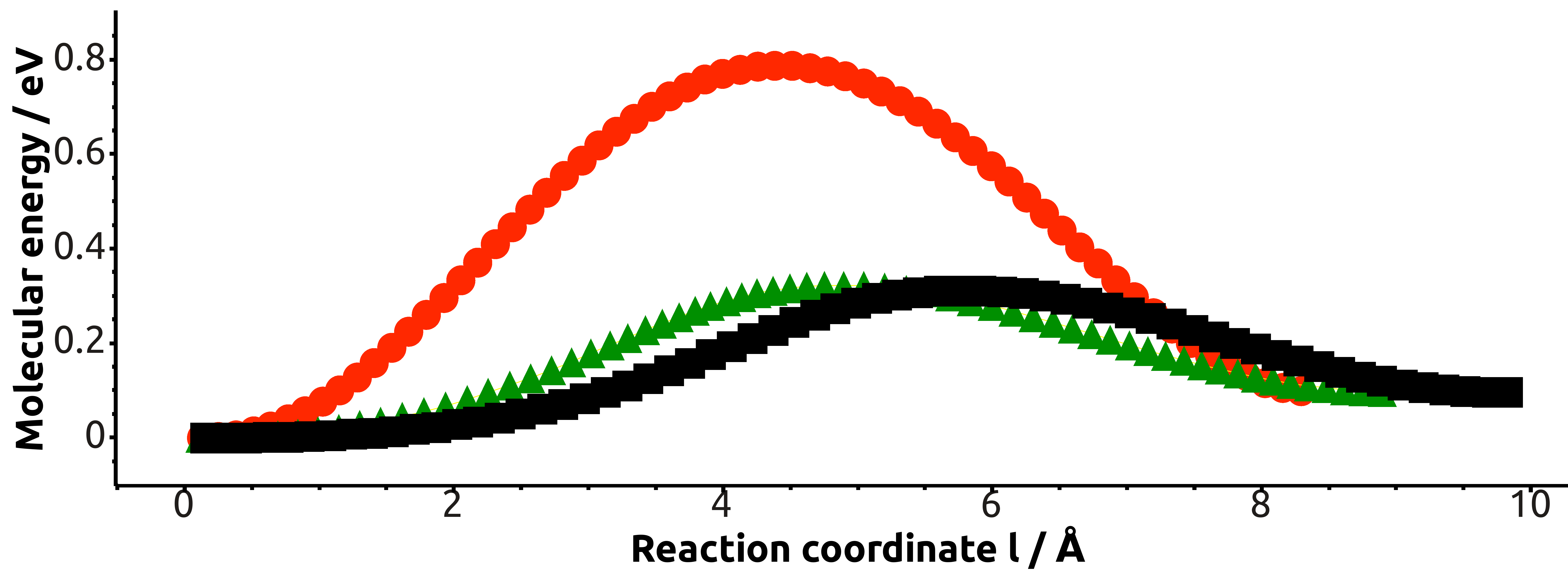}\\
  \caption{The evolution of the molecular energy  along the  reaction coordinate $l$ of the  $C7_{eq}\rightarrow C7_{ax}$ transition of alanine di-peptide. 
 The circles denote the classical dominant reaction pathway, obtained minimizing the HJ function (\ref{SHJ}),  the squares represent 
the minimum-energy path obtained from (\ref{SMEP}) and the triangles the dominant pathway with quantum corrections, obtained including the quantum component of the effective potential in the HJ action. }
 \label{Fig6}
 \end{figure}

\begin{figure}[t]
 \includegraphics[width=8.8 cm]{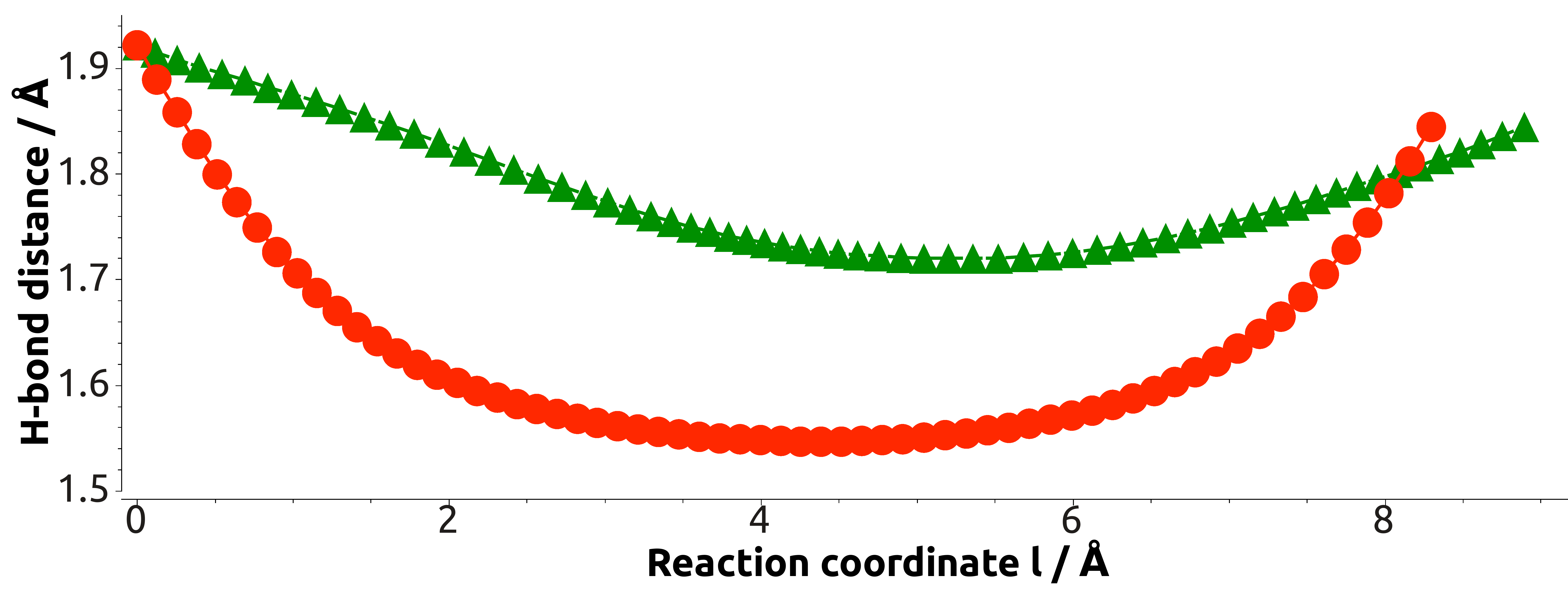}
  \caption{The evolution of the distance between the H18 and the O6 atoms,  along the reaction coordinate $l$ of the $C7_{eq}\rightarrow C7_{ax}$ transition of alanine di-peptide. 
  The circles denote the classical dominant reaction pathway, obtained minimizing
the HJ function (\ref{SHJ}) and the triangles the dominant pathway with quantum corrections, obtained including the quantum component of the effective potential in the HJ action. }
 \label{Fig7}
 \end{figure}

In Fig.~\ref{Fig3} we plot the dominant paths obtained in the DRP approach with and without quantum corrections, together with the minimum-energy path.
In the background we show the free-energy  as a function of the dihedral angles, evaluated  by means of an all-atom classical
 meta-dynamics simulation\cite{metadynamics, plumed}, using the same force field.   
In Fig.~\ref{Fig6} we present the evolution of the molecular energy along the reaction path,
while in Fig.~\ref{Fig7}, we report the evolution of the distance between the H18 and the O6 atoms, which are involved in  a hydrogen bond.
The entire set of calculations required in total about 700 hours on 2.2 GHz processors.

Some comments on these results are in order. First of all, we observe that the classical dominant pathway crosses the barrier in a region
 in which the molecular potential energy is about twice as large than at the saddle point, which is visited by the minimum-energy path, see Fig. \ref{Fig6}. 
We emphasize the fact that the dominant reaction pathways are expected to describe genuinely \emph{non-equilibrium} transitions. In general, such paths do not need
 to cross the barrier precisely at the saddle-point. On the other hand, it is important to check that such a large effect is not an artifact of the calculation. For example, problems may emerge
 if the effective energy parameter $E_{eff}$ was chosen very large. In this case, the total transition time would be very small and the calculation would lead information about the dynamics of 
 ultra-fast transitions. In addition, problems may emerge if the path space was
 not sufficiently explored during the minimization procedure. 
 In order to check the numerical reliability of our results, we have computed the classical dominant pathway, starting from a 
 path which crosses the  barrier at the saddle point, with an effective energy only $1\%$ larger  than the maximum value of $|V_{eff}[x]|$ along the initial path. 
 After the minimization of the  HJ action, we recovered the same result for the classical dominant path shown in Fig. \ref{Fig3}, and a very similar transition time. This result makes us
  confident that the dominant paths are independent on the choice of the initial trial path and are not appreciably dependent on the specific choice of the effective energy. 
  
A second important result of our calculation is that the quantum effects on the structure of the dominant reaction pathways 
are clearly visible. Even though all the three paths are qualitatively similar, the energy difference which is overcome by the most probable 
reaction pathway in the presence of quantum fluctuations is about $50\%$ smaller than the one in the classical case (see Fig.~\ref{Fig6}). 
Such a large difference arises because the
molecular energy surface in the transition region is quite steep and the  quantum fluctuations of hydrogen atoms are quite large. 
Note that the (classical) free energy differences in the same region are much smaller, due to a relatively high entropic contribution, associated e.g. to the rotation of the other
 dihedral angles. In addition, the distance between the atoms
 involved in the hydrogen bond is always about $0.2$~\AA\ larger in the quantum than in the classical dominant path (see Fig.~\ref{Fig7}). This suggests that the energy difference 
 overcome by the  classical dominant path is larger, because the O and H atoms  get closer, hence increasing their van der Waals repulsion.  
  
\begin{figure}[t]
 \includegraphics[width=8 cm]{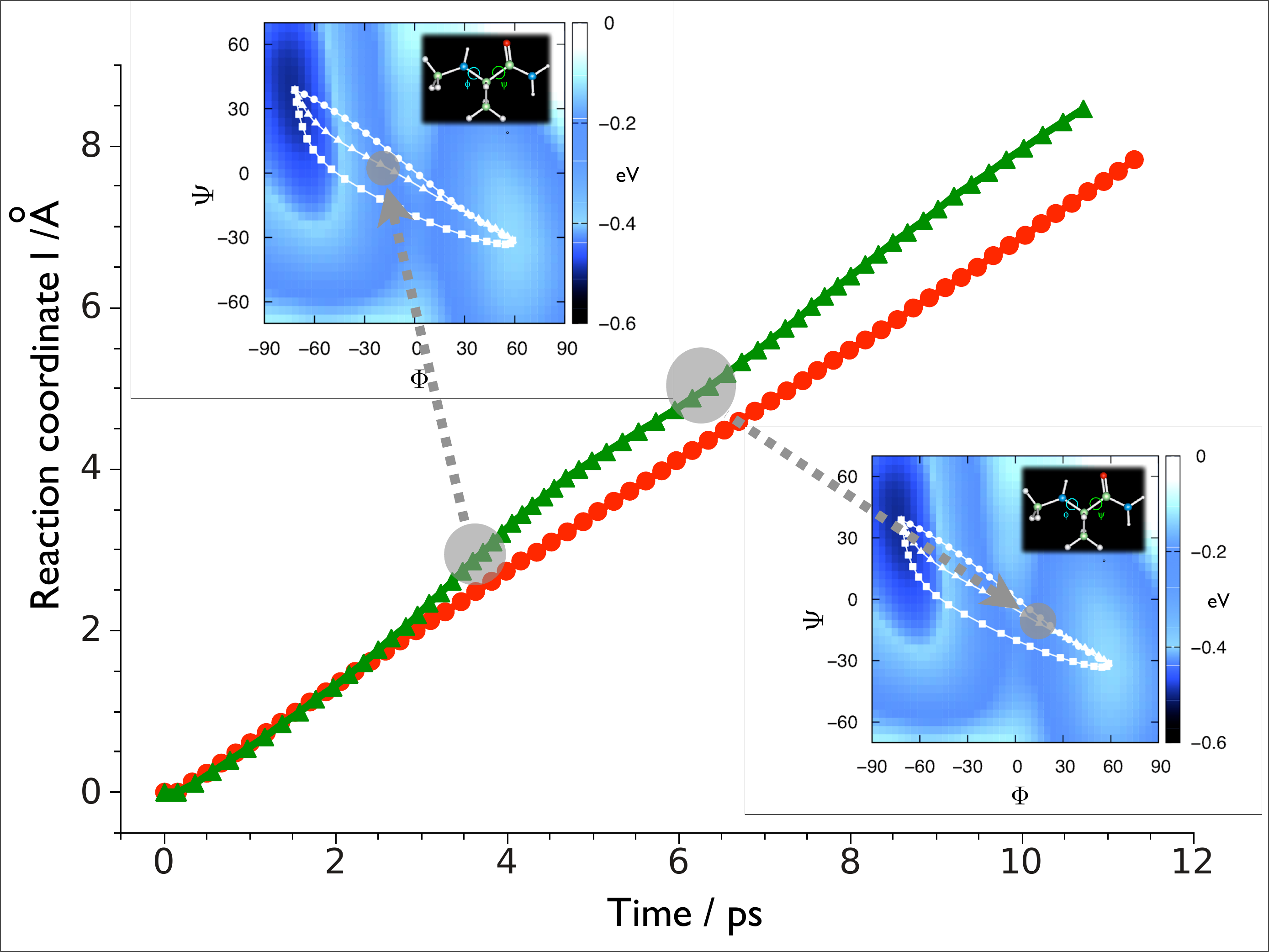}
  \caption{The time evolution of the dipeptide, in the  $C7_{eq}\rightarrow C7_{ax}$ transition of alanine di-peptide. 
 The circles denote the classical dominant reaction pathway, obtained minimizing the HJ function (\ref{SHJ}), the triangles the dominant pathway with quantum corrections, 
 obtained including the quantum component of the effective potential in the HJ action. In the inserts we show the position of some frames along the reaction path. }
 \label{Fig8}
 \end{figure}

In Fig. \ref{Fig8} we compare the time evolution of the system in the classical and quantum calculations, i.e.  we plot the time at which each value of the reaction coordinate 
is visited along the transition. First of all, we note that in both calculations the
most probable transition lasts about $15$~ps, i.e. a time much longer than the $0.2$~ps time scale below which the overdamped approximation is no longer appropriate.
It is also interesting to compare the velocity of the classical (circles) and quantum (triangles) dominant transition, which is represented by the slope of the curves $l(t)$. We see that while the velocity along
the classical path is essentially constant throughout the entire reaction, the quantum dominant path accelerates after about $3$~ps after about $4$~ps, i.e. in the region of high force before and after the saddle. 
 \section{Conclusions}
\label{conclusions}

In this work we have introduced a formalism which allows to investigate  at the semi-classical level the role of quantum fluctuations of atomic nuclei in the real-time dynamics of non-equilibrium molecular transitions.
Unlike other method which are more suited for rate calculations \cite{instanton1, instanton2, instanton3} or exploring the short-time dynamics inside a thermodynamical state \cite{centroid}, the present DRP approach 
is particularly efficient in investigating the real time dynamics as the system is crossing the free-energy barrier. The computational efficiency of the method makes it possible to study reactions involving large molecules, such as e.g. peptide chains. 

From an analysis of the quantum corrections to the Langevin equation we have shown that the quantum stochastic dynamics in the vicinity of the saddles with negative Hessian trace is qualitatively similar to a classical 
stochastic dynamics at a lower temperature. Conversely, in the vicinity of saddles with positive Hessian trace, or in the potential wells, quantum fluctuations can be interpreted as effectively raising the temperature.
 
We have shown that, in the test case of H$_2$ dissociation on copper,  the quantum corrections obtained in the DRP formalism away from equilibrium qualitatively agree with those 
obtained in equilibrium conditions, using the instanton method. 
 
We have applied the DRP formalism  to the  study of the $C7_{eq}\rightarrow C7_{ax}$ transition of alanine dipeptide which represents a prototypical example of biomolecular transition involving a 
hydrogen bond. We have found that in this reaction the inclusion of   
quantum fluctuations can significantly modify the reaction path with respect to a classical calculation. 

We conclude this work by discussing  possible limitations of the present approach. In general, we expect the DRP method to become inefficient in the following scenarios:
\begin{itemize}
\item
For each different boundary condition (\ref{bc}) there exists a large number of local minima of the HJ functional, all with comparable statistical weight, $\exp(-S_{HJ})$. 
\item
 The reaction mechanism depends very strongly on the initial 
configuration ${\bf X}_i$ and the reactant space is large. In this case, a very large number of reaction pathways would be needed in order to fully characterize the transition. By contrast,  
any method which provides only a relatively small number of them  would carry insufficient information (note that this limitation applies also to MD simulations). In this case, one must rely on a 
description based on the projection on a small  set of reaction coordinates.
\item
The fluctuations around each of the different dominant paths are very large and the regions visited by the fluctuations associated to different dominant paths significantly overlap. 
In this case,  the very notion of 
dominant pathway looses its significance. On the other hand, if such fluctuations are relatively small, their contribution can be systematically included through 
a perturbative expansion in the thermal energy  $k_B T$, using the method recently developed in Ref.~\cite{DRPrate}.  
\end{itemize}
In order to assess how such potential  limitations affect the applicability of the DRP method to realistic molecular transitions, several comparative tests were performed, based on the comparison 
against the result obtained by MD simulations~\cite{DRPtest2, DRPtest3}. These studies have shown that the DRP approach 
yields the correct description of the non-equilibrium dynamics of complex macromolecular transitions,  such as protein folding.  On the other hand, we emphasize that the semi-classical approach presented here works in conditions in which the quantum effects provide at most small 
corrections to  \emph{thermally activated pathways}, hence in the presence of dissipative dynamics. It is not applicable to investigating the non-dissipative tunneling and in 
general the dynamics in the deeply quantum regime.

\acknowledgments
The authors are grateful to Prof. Hannes J{\'onnson} for providing the
computer code calculating the potential described in Ref.~ \cite{instanton1.75} and to 
F. Pederiva for many important discussions.

 All the authors are members  of the Interdisciplinary 
Laboratory for Computational Sciences (LISC), a joint venture between  University of Trento and Fondazione Bruno Kessler.
S. a Beccara is supported by the \emph{Provincia Autonoma di Trento}, through the AuroraScience project. 

 Simulations were performed on
 the WIGLAF cluster at the Physics  
Department  of the University of  Trento and on the AURORA supercomputer at the LISC.

\appendix

\section{Multi-dimensional Generalization of the Quantum Smoluchowski Equation}
\label{generalization}

In this appendix, we provide the generalization of the QSE (\ref{QSE1D}) to a physical system consisting of $3~N$ degrees of freedom (e.g. the atomic coordinates of a molecule). 
We begin by observing that the \emph{naive} substitution $\frac{d}{dx}\to \nabla$  in Eq. (\ref{QSE1D}) does not represent the correct generalization, as it does not lead to the correct
  equilibrium distribution $P_{eq}(x)$,  describing the thermodynamical limit in the semi-classical regime. 
In order to obtain the correct multi-dimensional generalization of the QSE (\ref{QSE1D}) one can use the same path integral approach adopted in \cite{QSE}. Since such a procedure 
is quite lengthy and technically rather involved, here we present an alternative, albeit slightly less rigorous, derivation leading to the same result.  

The conservation of the number of particles implies a continuity equation for the probability, i.e.
\be
\label{cont}
\partial_t P({\bf X},t) = {\bf \nabla} \cdot {\bf J}({\bf X},t) = \sum_{i=1}^N \vec{\nabla}_i~\cdot~\vec{j}_i({\bf X},t),
\ee
where ${\bf J}({\bf X},t) \equiv (\vec{j}_1({\bf X},t), \vec{j}_2({\bf X},t), \ldots, \vec{j}_N({\bf X},t))$ is the probability current. 

Without loss of generality, each single-particle component of the current $\vec{j}_i({\bf X},t)$ can be defined in terms of a set of in-so-far unspecified functions
$\vec{\xi}_1^i({\bf X})$, and $\xi^i_2({\bf X})$:
\be
\label{ji}
\vec{j}^i({\bf X}, t) &\equiv& D^i_0~\left(\beta \vec{\nabla}^i U({\bf X}) + \vec{\xi}_1^i({\bf X}) \right)  P({\bf X},t)\nonumber\\
&+& \vec{\nabla}^i \left[D_0^i~( 1 + \xi_2^i({\bf X})) P({\bf X},t)\right],
\ee
where $D_0^i = 1/(m_i \beta \gamma)$ are the classical diffusion constants. Such a definition assures that in the limit $\vec{\xi}_1^i({\bf X}), \xi_2^i({\bf X}) \to 0$, one recovers the classical
Smoluchowski equation. The functions
\be
\vec{D}_1^i({\bf X}) &=& D^i_0~\left(\beta \vec{\nabla}^i U({\bf X}) + \vec{\xi}_1^i({\bf X}) \right) \\
D^i_2({\bf X}) &=& D_0^i~( 1 + \xi_2^i({\bf X}))
\ee
are the multi-dimensional generalization of the Moyal coefficients discussed in~\cite{QSE}. 

The unknown functions $\vec{\xi}_1^i({\bf X})$ and  $\xi_2^i({\bf X})$ can be determined by requiring that the continuity Eq. (\ref{cont}) must yield the correct thermodynamics, i.e. that its stationary 
solution  coincides with the well-known semi-classical expression of the Boltzmann's weight, 
\be
\label{Peqa}
P_{eq}({\bf X}) &=& \exp\left(-\beta U({\bf X}) \right)~\left( 1- L_1({\bf X}) + L_2({\bf X})\right),\qquad
\ee
with
\be
L_1({\bf X}) &\equiv& \beta \sum_{k=1}^N \lambda_k  ~\vec{\nabla}_k^2 U({\bf X})\\
L_2({\bf X}) &\equiv& \frac{\beta^2}{2} \sum_{k=1}^N \lambda_k~ |\vec{\nabla}_k U({\bf X})|^2
\ee
Using Eq. (\ref{ji}) and Eq. (\ref{Peqa}) and imposing the condition of vanishing current at thermal equilibrium, $\lim_{t\to \infty} \vec{j}({\bf X}, t)=~0$, up to leading order in the quantum 
expansion parameters, we obtain 
\be
&& \vec{\xi}_1^i({\bf X}) + \vec{\nabla}^i \xi_2^i({\bf X})  +  \vec{\nabla}^i L_2({\bf X}) \nonumber\\
&&- \vec{\nabla}^i L_1({\bf X})  - \xi^i_2({\bf X}) \beta \vec{\nabla}^i U({\bf X})=0 
\ee
The consistency with one-dimensional result (\ref{QSE1D}) implies
\be
\label{xi2}
\xi_2^i({\bf X}) = L_1({\bf X}).
\ee
Hence, 
\be
\label{xi1}
\vec{\xi}^i_1({\bf X}) &=& L_2({\bf X}) \beta \vec{\nabla}^i U({\bf X}) - \vec{\nabla}^i L_2({\bf X}). 
\ee
The QSE obtained from Eq.s (\ref{cont}), (\ref{ji}), (\ref{xi2}) and (\ref{xi1}) is equivalent to Eq. (\ref{QSEq}),  to leading order in the quantum expansion parameters $\lambda_k$.  
However, the form (\ref{QSEq}) is usually preferred, as it guarantees the consistency with the 
second law of thermodynamics \cite{Hanggi}.

\section{Path Integral Representation of Quantum Langevin Dynamics}
\label{appendixB}

In this appendix, we show that the solution of the QSE (\ref{QSEq}) can be sampled by integrating an associated quantum Langevin equation, with a multiplicative noise. 
We also construct the path integral representation ({\ref{PI2}}), which is used to derive the quantum extension of the DRP formalism. 

Let us begin by considering a generic Langevin  equation with multiplicative noise, in the form
\be
\label{genL}
\dot{\vec{x}}_i =  \vec{f}_i({\bf X}) + g({\bf X}) \vec{\eta}_i(t), \qquad (i=1,\ldots, N)
\ee
where  $\vec{x}_i$ denotes the coordinates of the $i-$th particle, $\vec{\eta}_i(x)$ is a 3-dimensioanl stochastic force of unit variance. 
Such a stochastic differential equation generates a probability distribution which obeys  the generalized Smoluchowski equation ---see e.g. discussion in \cite{D(x)}--- 
\be
\label{GSE}
\frac{\partial}{\partial t} P({\bf X},t) &=& \sum_{i=1}^N \vec{\nabla}_i \left[ \left(- \vec{f}_i({\bf X}) - \alpha g({\bf X}) \vec{\nabla}_i g({\bf X}) \right)~ P({\bf X},t)\right.\nonumber\\
 &+&  \left.\frac{1}{2}\vec{\nabla}_i~\left(g^2({\bf X})~P({\bf X},t)\right)~\right] 
\ee
The real parameter $0\le \alpha\le 1$ specifies the stochastic calculus adopted to define the differential Eq.~(\ref{genL}). In particular, $\alpha=0$ ($\alpha=1/2$) corresponds to the so-called 
Ito (Stratonovich) calculus. 

We now want to derive a Langevin equation in the form (\ref{genL}) which generates a probability density obeying the QSE (\ref{QSEq}). To this end, it is important to emphasize that QSE is an ordinary partial differential equation, hence it is uniquely defined in the standard (e.g. Riemann) calculus. Hence,  for every choice of stochastic calculus $\alpha$ there is in general a different Langevin equations, associated to the same physical Smoluchowski Eq.~(\ref{QSEq}).
This can be obtained by finding the functions $g_i({\bf X})$ and the vector fields $\vec{f}_i({\bf X})$ such that (\ref{GSE})
Smoluchowski equation coincides with the QSE (\ref{QSEq}), to order $\lambda$ accuracy. Such a request leads to
 \be
 \label{match1}
 g_i({\bf X}) &=& \sqrt{2 D_i} ~\left(1+\frac{1}{2} L_1({\bf X})~\right)\\
 \label{match2}
\vec{f}_i({\bf X}) &=& - D_i \beta \vec{\nabla}_i U({\bf X}) + D_i \beta \vec{Q}_i({\bf X}),
 \ee
where $\vec{Q}_i({\bf X})$ is  a "quantum force" whose definition depends on the calculus adopted and reads 
 \be
\vec{Q}_i({\bf X}) &=& \frac{1}{\beta} \vec{\nabla}_i L_2({\bf X}) \nonumber\\
&-& L_1({\bf X}) \vec{\nabla}_i U({\bf X}) - \frac{\alpha}{\beta} \vec{\nabla}_i L_1({\bf X}). 
 \ee

Also the path integral representation of the solution of the quantum Smoluchowski Eq. (\ref{QSEq}) 
depends on the choice of the stochastic calculus and reads~\cite{D(x)}
 \be
\label{PIgen}
&& P({\bf X},t|{\bf X}_i) = \int_{{\bf X}_i}^{\bf X} \bar{\mathcal{D}}{\bf X}~\exp\left\{-~\int_0^t d\tau 
\sum_{i=1}^N\left[ \frac{1}{2 g_i^2({\bf X})}\right.\right.  \nonumber\\
&&\left. \left. \cdot \left| \vec{\dot x}_i -  \vec{f}_i({\bf X}) + \alpha g({\bf X}) \vec{\nabla}_i g_i({\bf X}) 
 \right|^2 + \alpha \vec{\nabla}_i\cdot \vec{f}_i({\bf X})\right]\right\},\nonumber\\
 \ee 
 where $g_i({\bf X})$ and $\vec{f}_i({\bf X})$ are given by Eq.s (\ref{match1}) and (\ref{match2}), and 
 the modified  Wiener measure $\bar{\mathcal{D}}{\bf X}$  depends on the configuration and reads 
 \be
 \label{meas}
\bar{ \mathcal{D}}{\bf X} = \lim_{N_t\to \infty}~\prod_{l=1}^{N_t}\prod_{i=1}^N~\frac{d \vec{x}_i(l) }
{\left[4 D_i \pi \Delta t~ (1+ \lambda_i \beta \nabla_i^2 U[{\bf X}(l)] )\right]^{3/2}},\nonumber\\
\ee
where $\Delta t = t/N_t$.
Plugging (\ref{match1})-(\ref{match2}) into Eq. (\ref{PIgen}) and expanding to leading order in the $\lambda_i$ we obtain, after some tedious but rather straightforward calculations
\be
\label{PIQ}
&&P({\bf X},t|{\bf X}_i) = \int \mathcal{D}{\bf X}~
\exp \left\{-\int_0^t d\tau \sum_i \left[ \frac{\vec{\dot x}_i}{2}~ \cdot\left( \beta \vec{\nabla}_i U({\bf X}) \right. \right. \right. \nonumber\\
&&\left. \left. \left. - \beta \vec{Q}^i - \alpha \vec{\nabla}_i L_1({\bf X})\right)\right]\right\} \exp\left\{-\int_0^t d\tau \sum_i \left[
 \frac{\vec{\dot x}_i^2}{4 D_i} \right.\right.~ \nonumber\\%
&&\left.\left. + \frac{D_i \beta^2}{4} \Large(|\vec{\nabla}_i U({\bf X})|^2    
 - \frac{2}{\beta} \vec{\nabla}_i^2 U({\bf X})\Large) + \alpha \beta D_i \vec{\nabla}_i \cdot
  \vec{Q}_i({\bf X})   \right.\right. \nonumber\\
&&\left.\left. - \frac{D_i}{4} \beta^2 |\vec{\nabla}_i U({\bf X})|^2 L_1({\bf X})+ \alpha \beta D_i  \vec{\nabla}_i L_1({\bf X}) \cdot \vec{\nabla}_i U({\bf X})  \right] \right\}.\nonumber\\
\ee
Note that the path integral now contains the standard Wiener measure 
\be
 \mathcal{D}{\bf X} = \lim_{N_t\to \infty}~\prod_{l=1}^{N_t}\prod_{i=1}^N~\frac{d \vec{x}_i(l) }
{(4 D_i \pi \Delta t )^{3/2}}.
\ee

The first exponent can be taken out of the path integral since it does not affect the statistical weight of the diffusive paths.
To see this, we introduce a scalar function $W(x)$ which is  defined as the formal solution of the partial differential equation
\be
\label{W}
\vec{\nabla}_i W({\bf X}) = \left(-\vec{\nabla}_i U({\bf X}) + \beta Q_i - \alpha \vec{\nabla}_i L_1({\bf X})\right)
\ee
With such a definition,  the first exponent in Eq. (\ref{PIQ}) is written as an exact differential form,
\be
\label{Nform}
e^{-\int_0^t d\tau ~\sum_i \frac{\vec{\dot{x_i}}}{2}\cdot~\vec{\beta \nabla}_i W({\bf X})} &=& e^{-\frac\beta2 \int_0^t d\tau ~\frac{d}{dt} W({\bf X})}\nonumber\\
&=& e^{-\frac{\beta}{2}(W({\bf X}_f)- W({\bf X}_i))}\nonumber\\
&\equiv& \mathcal{N}({\bf X}_f, {\bf X}_i) 
\ee
which depends only on the end-points and not on the path.   If we now specialize on the Stratonivich calculus we obtain Eq. (\ref{PI2}).


\begin{thebibliography}{99}
\bibitem{QMM} A.R. Leach,  "Molecular modeling: principle and applications" (2nd ed.) Pearson Education (Harlow, England), 2001.
\bibitem{Landau} L. D. Landau and E. M. Lifshitz, "Statistical Physics Part 1" (1980, 3rd Ed. ). Butterworth-Heinemann (Oxford).
\bibitem{review} H. Grabert, P. Schramm and G.-L. Ingold, Phys. Rep. {\bf 168}, 115 (1988). 
\bibitem{QSE} J. Ankerhold, P. Pechukas and H. Grabert, Phys. Rev. Lett. {\bf 87}, 086802 (2001). J. Ankerhold and H. Grabert, Phys. Rev. Lett. {\bf 101}, 119903 (2008) (Erratum). 
J. Ankerhold, Phys. Rev. {\bf E64}, 060102 (2001). 
\bibitem{Hanggi} L. Machura, M. Kostur, P. H\"anggi, P. Talkner and J. Luczka, Phys. Rev. {\bf E 70}, 031107 (2004). 
\bibitem{Coffey}  W. T. Coffey, Y. P. Kalmykov, S. V. Titov, and B. P. Mulligan, J. Phys. {\bf A 40}, F91(2007).  W. T. Coffey, Y. P. Kalmykov, S. V. Titov, AND L. Cleary, Phys. Rev. {\bf E 78} 031114 (2008).   
\bibitem{centroid} S. Jang and G.A. Voth, J. Chem. Phys. {\bf 111}, 2371 (1999). 
\bibitem{instanton1} W. H. Miller, J. Chem. Phys. {\bf 62}, 1899 (1974).
\bibitem{instanton1.5} G. Mills, and H. J{\'o}nsson, Phys. Rev. Lett. {\bf 72}, 1124 (1994).
\bibitem{instanton1.75} G. Mills, H. J{\'o}nsson and G .K.~ Schenter, Surf. Sci. {\bf 324}, 305 (1995).
\bibitem{instanton2} G. Mills, G.K. Schenter, D. E. Makarov, and H. J{\'o}nsson, Chem. Phys. Lett. {\bf 278}, 91 (1997).
\bibitem{instanton3} S. Althorpe J. Chem. Phys. {\bf 131}, 214106 (2009)
\bibitem{textbook} U. Weiss, "Quantum Dissipative Systems" 3rd Ed. (World Scientific, Singapore 2008). 
\bibitem{phivalue} A. Matouschek, J. T. Kellis Jr., L. Serrano and A. R. Fersht, Nature {\bf 340}, 122 (1989). 
\bibitem{singlemoleculeexp} C. Cecconi, E. Shank, C. Bustamante and S. Marquesee, Science {\bf 309}, 2057 (2005). E. A. Shank, C. Cecconi, J. W. Dill, S. Marquesee and C. Bustamante, Nature {\bf 465} 637 (2010).  
 \bibitem{DRPtheory1} P. Faccioli, M. Sega, F. Pederiva  and H. Orland, Phys. Rev. Lett. {\bf 97}, 108101 (2006).   
 \bibitem{DRPtheory2} E. Autieri, P. Faccioli, M. Sega, F. Pederiva  and H. Orland,  J. Chem Phys. {\bf 130}, 064106 (2009).  
\bibitem{DRPrate} G.Mazzola, S. a Beccara, P.Faccioli, and H. Orland,  J. Chem. Phys. {\bf 134}, 164109 (2011).
\bibitem{Elber0} R. Elber,  and D. Shalloway, J. Chem. Phys. {\bf 112}, 5539 (2000).
\bibitem{DRPtest1}  M. Sega, P. Faccioli, F. Pederiva, G. Garberoglio and H. Orland, Phys. Rev. Lett. {\bf 99}, 118102 (2007).
\bibitem{DRPtest2} P. Faccioli, J. Phys. Chem. {\bf B112}, 137560 (2008).
\bibitem{DRPtest3}  P. Faccioli, A. Lonardi and H. Orland,  J. Chem. Phys. {\bf 133}, 045104 (2010).
\bibitem{QDRP1} S. a Beccara, G. Garberoglio, P. Faccioli and F. Pederiva, J. Chem. Phys. {\bf 132}, 111102 (2010).
\bibitem{QDRP2} S. a Beccara, P. Faccioli, M. Sega, G. Garberoglio, F. Pederiva and H. Orland, J. Chem. Phys. {\bf 134}, 024501 (2011).
\bibitem{D(x)} A. W. C. Lau and T. C. Lubensky, Phys. Rev. {\bf E 76} , 011123 (2007).
\bibitem{AMBER}  D.A. Case, T.E. Cheatham III, T. Darden, H. Gohlke, R. Luo, K.M. Merz, Jr., A. Onufriev, C. Simmerling, B. Wang and R. Woods,  J. Computat. Chem. {\bf 26}, 1668 
(2005).
\bibitem{FIRE} E. Bitzek, P. Koskinen,  F. G\"ahler, M. Moseler, and P. Gumbsch,  Phys. Rev. Lett. {\bf 97}, 170201 (2006).
\bibitem{NEB} G. Henkelmann and H. J\'onsson,  J. Chem. Phys., {\bf 113},  9978 (2000). 
\bibitem{metadynamics} A. Laio and M. Parrinello, Proc. Natl. Acad. Sci. U.S.A. 99, 12562 (2002)
\bibitem{plumed} M. Bonomi, \emph{et. al},  Comp. Phys. Comm. {\bf 180}, 1961 (2009).
\end{thebibliography}
\end{document}